\documentclass{llncs}
\usepackage{graphicx}
\usepackage{amssymb}
\usepackage{amsmath}
\usepackage{mathtools}

\newtheorem{probl}{Problem}

\newcommand{\tssl}[0]{\textstyle\sum\limits}

\begin{document}
\title{A Natural Quadratic Approach to the Generalized Graph Layering Problem}
%
%
\author{Sven Mallach\inst{1}}
\authorrunning{S. Mallach}
%
\institute {
Dept.\@ of Mathematics and Computer Science, University of Cologne, Germany\\
  \email{mallach@informatik.uni-koeln.de}
}
\maketitle              

\setcounter{footnote}{0}
\begin{abstract}
We propose a new exact approach to the generalized
graph layering problem that is based on a particular
quadratic assignment formulation. It expresses,
in a natural way, the associated layout restrictions
and several possible objectives, such as a minimum total
arc length, minimum number of reversed arcs, and minimum
width, or the adaptation to a specific drawing area.
Our computational experiments show a competitive
performance compared to prior exact models.
\keywords{Graph Drawing \and Layering \and Integer Programming}
\end{abstract}
\section{Introduction}\label{s:Intro}

Hierarchical graph drawing is an indispensable tool to
automate the cleaned-up illustration of e.g.~flow
diagrams or data dependency representations.
Here, the dominant methodological framework studied in
research and implemented in software is the
one proposed by Sugiyama et al.~\cite{SugiyamaTT81}.
It involves four successive and interdependent steps for
cycle removal,
vertex layering,
crossing minimization,
and,
finally, horizontal coordinate assignment and arc~routing.

Classically, the workflow
is carried out by solving the feedback arc set
problem, i.e., reversing (a minimum number of)
arcs, in the first step such that all arcs have
a common direction during the others. Then, for the final
drawing, original directions are restored.
As a result, the height of a graph's layering is bounded
from below by the total height or number of vertices on a
longest path after the first step.
In particular, a poor aspect
ratio of the final drawing may thus
be inevitable from the very beginning. Also, the number and
placement of dummy vertices, usually introduced if an
arc spans a layer to facilitate the other steps and to
more accurately account for the width, is strongly affected.

\begin{figure}[h]
\centering
\includegraphics[scale=0.4]{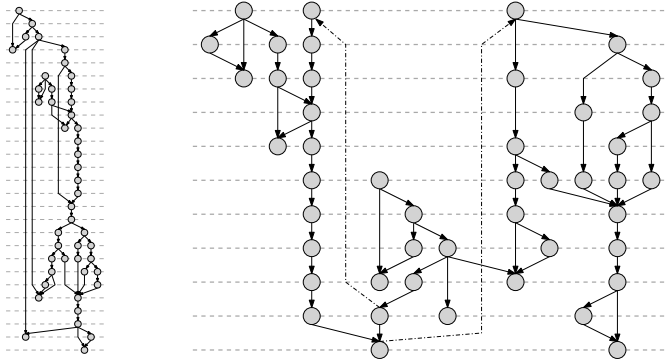}
    \caption{Two layered drawings of the same directed acyclic graph. On the
    left a classic one, i.e., adhering to its longest path with all arcs
    pointing downwards, and on the right with a better
    aspect ratio achieved by reversing only two arcs (drawn dash-dotted).}
\label{fig:motivation}
\end{figure}

This motivates the recently studied integration of the first two
steps~\cite{JabrayilovMMR+16,RueeggESvH15,RueeggESvH16,JGAA-441}.
Here, the central idea is to identify (a small number of) suitable
arcs to be drawn reverse to the intended hierarchical direction
such that this enables a layout that is \emph{two-dimensionally}
compact, possibly meets other common aesthetic criteria such as
a minimum total arc length, or even adapts to a drawing area of
a certain aspect ratio. Fig.~\ref{fig:motivation} exemplifies the
potential effects on aesthetics and readability. Moreover, previous
experimental studies in the referenced articles
show that significant improvements in terms of the drawing area or aspect
ratio are frequent when optimizing the layering
with respect to
the corresponding~objectives.

In this paper, we present a new exact approach to integrate
vertex layering, the feedback arc set problem, and width or
drawing area optimization. Its most appealing novelty is
that it adheres to the quadratic nature of the problem
rather than avoiding it. This quadratic
nature does not only exist geometrically. Even when
optimizing only for one direction (e.g.\ width),
several aspects of a layered drawing depend on
\emph{conjunctive} decisions. For instance,
the questions whether an arc is reversed,
or whether it causes a dummy vertex on a particular layer,
depend on the layer assignments of both of the arc's end
vertices at the same~time.

Our approach allows to express such conditions
that depend on two simultaneous decisions, and
thus all the present generalizations of
the classical graph layering problem and their
objective functions, in a natural and intuitive~way.
It is based on a quadratic assignment
problem (QAP) formulated and solved as a mixed-integer
program (MIP). As our computational results show, it
can compete with the currently best known, but
less intuitive MIP formulations. This is
surprising since the QAP is considered to be
among the hardest $\mathcal{NP}$-hard
optimization problems. However, the graph layering problem
poses a particularly well-suited special case: Our model
does not require artificial constructions to model
common drawing restrictions and objectives as linear
expressions but profits from a sophisticated linearization
technique, is compact in the number of constraints, and
comparably insensitive to a graph's density.
Ideally, our drawn links to the QAP inspire also new models for related layout styles
or heuristic approaches to tackle larger instances or to support
interactive user applications.

The paper is organized as follows. In Sect.~\ref{s:Landscape}, we
give a quick survey of the state of the art generalizations
of the classical graph layering problem.
The major existing exact approaches to solve these are
highlighted in Sect.~\ref{s:Exact}.
We then present our quadratic formulation
of the most generalized graph layering problem variants
in Sect.~\ref{s:Nature}. Finally, we report in Sect.~\ref{s:Exp} on our
computational evaluation, and close with a conclusion in Sect.~\ref{s:Concl}.

\section{A Landscape of Graph Layering Problems}\label{s:Landscape}

A \emph{layering} $L$ of a directed graph $G=(V,A)$ with vertex set $V$
and arc set $A$ is a mapping $L: V \rightarrow \mathbb{N}^+$ that
assigns each vertex $v \in V$ a unique \emph{layer} index $L(v)$.
Classically, and presuming that $G$ is acyclic, a layering $L$ is considered
\emph{feasible} if $L(v) - L(u) \ge 1$ holds for all $uv \in A$.
In $1993$, the following associated problem was introduced
and shown to be polynomial time solvable by Gansner et al.~\cite{GansnerKNV93}:
\begin{probl}{Directed Layering Problem (DLP).}\label{pr:DLP}
Let $G=(V,A)$ be a directed acyclic graph. Find a feasible layering $L$ of $G$
minimizing the total arc length
\[\sum_{uv \in A} \big(L(v) - L(u)\big).\]
\end{probl}

Since, in a final layered drawing of non-acyclic graphs, the presence
of reversed arcs is inevitable, and to overcome the limitations
mentioned in the introduction (as well for acyclic graphs), a straightforward generalization of DLP
discussed by R\"uegg et al.~\cite{RueeggESvH15,RueeggESvH16}, is to
integrate arc reversals into the layering phase, and thus to consider a
layering $L$ feasible if $L(v) \neq L(u)$, i.e., $\lvert L(v) - L(u) \rvert \ge 1$,
holds for all $uv \in A$.
This gives rise to a second objective besides edge length minimization,
namely the minimization of the number of reversed arcs. The trade off
between both goals may be addressed by introducing respective
weights $\omega_{len}$ and $\omega_{rev}$ into the objective function.
The resulting optimization problem is then:
\begin{probl}{Generalized Layering Problem (GLP).}\label{pr:GLP}
Let $G=(V,A)$ be a directed graph. Find a feasible layering $L$ of $G$
minimizing
\[\omega_{len}\; \Big( \sum_{uv \in A} \lvert L(v) - L(u) \rvert \Big)\; +\; \omega_{rev}\; \lvert \{uv \in A \mid L(v) < L(u) \} \rvert.\]
\end{probl}
As opposed to DLP, GLP is $\mathcal{NP}$-hard, and it remains so even
if one of $\omega_{len}$ and $\omega_{rev}$ is~zero~\cite{RueeggESvH15}.
Both problems have also been combined with width minimization
which is worthwhile, even though the final drawing width is further
influenced by the horizontal coordinate assignment and arc
routing~\cite{RueeggESvH16}. In this context, the
\emph{estimated width} $\mathcal{W}$ of a layering
is given by the maximum number or maximum total width
of original and dummy vertices in any of its layers. With an
associated objective function weight $\omega_{wid}$, we have the two according problems:
\begin{probl}{Minimum Width Directed Layering Problem (DLP-W).}\label{pr:DLP-W}
Find a layering $L$ of a directed acyclic graph $G=(V,A)$ feasible for DLP
that minimizes
\[ \omega_{len}\; \Big(\sum_{uv \in A} (L(v) - L(u))\Big) + \omega_{wid}\; \mathcal{W}.\]
\end{probl}
\begin{probl}{Minimum Width Generalized Layering Problem (GLP-W)\footnote{This problem is called \emph{Compact Generalized Layering Problem (CGLP)} in~\cite{JabrayilovMMR+16,JGAA-441}
but renamed here to harmonize with the other variants.}.}\label{pr:GLP-W}
Find a layering $L$ of a directed graph $G=(V,A)$ feasible for GLP that minimizes
    \[\omega_{len}\; \big( \sum_{uv \in A} \lvert L(v) - L(u) \rvert \big)\; +\; \omega_{rev}\; \lvert \{uv \in A \mid L(v) < L(u) \} \rvert + \omega_{wid}\; \mathcal{W}.\]
\end{probl}

Here, it should be mentioned that DLP-W is equivalent to
the precedence-constrained multiprocessor scheduling problem
(when ignoring dummy vertices), and
thus as well $\mathcal{NP}$-hard~\cite{Ullman75}.

\begin{figure}[h]
\centering
\includegraphics[scale=0.4]{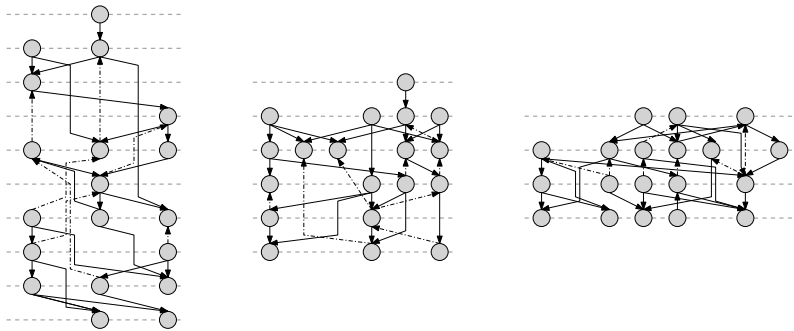}
    \caption{A directed graph drawn based on a layering created by solving
    GLP-MS with the ratio ${r_W:r_H}$ set to ${1:2}$ (left), ${1:1}$ (middle), and ${2:1}$ (right).
    The obtained optimal $\mathcal{W}:\mathcal{H}$ combinations are
    respectively ${5:10}$, ${6:6}$, and ${8:4}$.}
\label{fig:different_areas}
\end{figure}

Finally, R{\"u}egg et al.~propose to optimize a layering
with respect to
a target drawing area of width $r_W$ and height
$r_H$~\cite{JGAA-441}. Informally, a \lq best\rq\ such
drawing is considered one that can be maximally scaled
(\lq zoomed in\rq) until it exhausts one of the two
dimensions
(cf.\ Fig.~\ref{fig:motivation} and \ref{fig:different_areas}).

Formally, define $\mathcal{H} \coloneqq \max_{v \in V} L(v)$
to be the height
of a layering $L$.
This definition is suitable as we may assume w.l.o.g.~(and, if necessary, enforce a posteriori
for any given and feasible layering) that vertices are assigned to consecutive layers starting from
index~one.
The \emph{scaling factor} $\mathcal{S}$ to be maximized is then the
minimum of the ratios between the targeted and the actually used width
and height, respectively, i.e.,
$\mathcal{S} = \min \{ \frac{r_W}{\mathcal{W}}, \frac{r_H}{\mathcal{H}} \}$.
Adding once more a corresponding weight $\omega_{scl}$, the problem can be
expressed as follows:
\begin{probl}{Maximum Scale Generalized Layering Problem (GLP-MS).}\label{pr:GLP-MS}
Given a drawing area of
(normalized\footnote{
We remark that the parameters $r_W$ and $r_H$ used to
characterize the target drawing area
introduce undesired economies of scale since different values
representing the same aspect ratio lead to different numeric
maxima of~$\mathcal{S}$.
At the same time, it is not necessary to specify
the dimensions of the drawing area in \emph{absolute} values
as the goal is a best effort layering for the \emph{relative}
aspect ratio of the targeted area.
We thus propose to
normalize $r_W$ and $r_H$ to $\frac{r_W}{\min \{r_W, r_H\}}$
and $\frac{r_H}{\min \{r_W, r_H\}}$, respectively.})
width $r_W$ and (normalized)
height $r_H$, find a feasible layering $L$ of a directed graph $G=(V,A)$
that minimizes the expression
$$\omega_{len}\; \big( \sum_{uv \in A} \lvert L(v) - L(u)\rvert \big) + \omega_{rev}\; \lvert \{uv \in A \mid L(v) < L(u) \}\rvert - \omega_{scl}\; \mathcal{S}.$$
\end{probl}

A slight variation of GLP-MS as well proposed in~\cite{JGAA-441}
will be denoted GLP-MS$^*$: Here, instead of minimizing
${-} \omega_{scl}\; \mathcal{S}$, one minimizes
$\omega_{scl}\; \bar{\mathcal{S}}$
where $\bar{\mathcal{S}} \coloneqq \frac{1}{\mathcal{S}}$.

\section{Evolution of Exact Approaches to Graph Layering}\label{s:Exact}

To our best knowledge, all existing \emph{exact} methods are
based on integer programming, and can be coarsely classified based
on three different types of variables involved to model
the \emph{layering} of the vertices of a directed graph $G=(V,A)$:
\begin{itemize}
\item Assignment-based: Binary variables $x_{v,k}$ taking on value
one if
$L(v) = k$,
and taking on value zero otherwise.
\item Ordering-based: This refers to binary variables $y_{k,v}$ taking
on value one if $L(v) > k$, and taking on value zero otherwise (see, e.g., the appendix).
\item Direct: $L(v)$ is directly modeled as a general
\emph{integer} variable.
\end{itemize}

Table~\ref{tab:LayerMIPs} gives a quick overview of the formulations proposed so far.
Some of them are named
to ease their reference.
Of course,
GLP models can solve DLP (by setting $\omega_{rev} = \infty$),
width minimization may be turned off (by setting $\omega_{wid} = 0$), and models
to maximize the scaling factor can be used to minimize the width
(by setting $r_W = 1$ and $r_H = \infty$).
\begin{table}[h]
\centering
\begin{tabular}{ l    c    c    c   }
Problem    & Assignment-based & Ordering-based &  Direct \\\hline
 DLP        &           &                   & \cite{GansnerKNV93} \\
 DLP-W      & WHS \cite{HealyN02b}  &   &        \\
 GLP        &           &                     & \cite{RueeggESvH16}      \\
 GLP-W      & EXT \cite{JabrayilovMMR+16} & CGL-W \cite{JabrayilovMMR+16} & \\
 GLP-MS$^*$ &           & CGL-MS$^*$ \cite{JGAA-441} &
\end{tabular}
\label{tab:LayerMIPs}
\caption{An overview of pre-existing MIP models for the different layering problems.}
\end{table}

The direct approach is a perfect fit for the DLP since the resulting
problem can be solved in polynomial time by combinatorial algorithms
as discussed by Gansner et al.~\cite{GansnerKNV93}.
However, this property (more precisely, the underlying structure)
is lost when incorporating width constraints or arc
reversals, and the direct method becomes inferior to models with
binary variables in practice~\cite{RueeggESvH16,JabrayilovMMR+16}.

The first assignment-based formulation (WHS) was proposed by
Healy and Nikolov in~\cite{HealyN02b}. They considered width
\emph{constraints}, but their model may be easily altered to
solve DLP-W. Here, the breakthrough to computational tractability for small
and medium-sized graphs was to exploit the fixed direction of arcs.

Naturally, this could again not be preserved when it comes to GLP.
Rather, modeling whether an arc is reversed or causes a dummy vertex
comes then at the cost of additional variables and linearization
constraints. Moreover, as Jabrayilov et al.~\cite{JabrayilovMMR+16}
show, the corresponding assignment-based formulation (model EXT) for
GLP-W rendered inferior to the first ordering-based one (called CGL-W).
Consequently, in~\cite{JGAA-441}, R{\"u}egg et al.~used
the latter as a basis to design a MIP model
(that is, to the best of our knowledge, the only one so far)
for GLP-MS$^*$ and that we thus consistently refer to as CGL-MS$^*$.

However, as described in the following, a \emph{quadratic}
assignment-based approach re-enables the possibility to
express conditions regarding arc reversals and dummy
vertices intuitively, and without \emph{artificial}
linearization constraints.

\section{A Natural Quadratic Graph Layering Framework}\label{s:Nature}

\subsection{A Basic Quadratic Layer Assignment Model (QLA)}

Let $G=(V,A)$ be a directed graph, and let $Y$ be an
upper bound on the number of layers such that assignment
variables $x_{v,k}$ are to be introduced for all
$v \in V$ and all $k \in \{1, \dots, Y\}$. Consider
as well the variables $p_{u,k,v,\ell}$, for
all $uv \in A$ and all $k, \ell \in \{1, \dots, Y\}$, that
shall express the product $x_{u,k}\; x_{v,\ell}$. Then,
a basic formulation of the layering constraints for any of
the GLP problems (DLP as well, but this would permit to be
more restrictive)
can be expressed as follows:
\begin{align}
           & \tssl_{k=1}^{Y} x_{v,k}               &=\;   & 1       && \mbox{for all } v \in V \label{con:assignD} \\
	   & \tssl_{\ell=1}^{Y} p_{u,k,v,\ell}           &=\;   & x_{u,k} && \mbox{for all } uv \in A,\; k \in \{1, \dots, Y\} \label{con:linxuk} \\
	   & \tssl_{k=1}^{Y} p_{u,k,v,\ell}           &=\;   & x_{v,\ell} && \mbox{for all } uv \in A,\; \ell \in \{1, \dots, Y\} \label{con:linxvl} \\
	   & p_{u,k,v,k}                           &=\;   & 0       && \mbox{for all } uv \in A,\; k \in \{1, \dots, Y\} \label{con:noshareD} \\
           & x_{v,k}                               &\in\; & \{0,1\}    && \mbox{for all } v \in V,\; k \in \{1, \dots, Y\} \nonumber\\
           & p_{u,k,v,\ell}                           &\in\; & [0,1]      && \mbox{for all } uv \in A,\; k,\ell \in \{1, \dots, Y\}  \nonumber
\end{align}

Equations~(\ref{con:assignD}) let each vertex be assigned a unique layer.
Following the compact linearization approach~\cite{Mallach2018},
equations~(\ref{con:linxuk}) and~(\ref{con:linxvl}) establish that
variable $p_{u,k,v,\ell} = x_{u,k} \cdot x_{v,\ell}$ if the latter two take
binary values\footnote{
An intuitive interpretation for~(\ref{con:linxuk}) is: If $x_{u,k}$ is
zero, all products involving it must be zero as well. Conversely, if
$x_{u,k}$ is one, then exactly one of the $p_{u,k,v,\ell}$ on the left hand side (which
are \emph{all} the products of $x_{u,k}$ with different $x_{v,\ell}$ for some
fixed $v \in V$, $v \neq u$) need to be equal to one as well due to (\ref{con:assignD}).
Equations~(\ref{con:linxvl}) imply the same for the second factor
of any product variable $p_{u,k,v,\ell}$. Of course, these as well as
(\ref{con:linxuk}) and (\ref{con:linxvl}) could be avoided under employment
of an evolved \emph{non-linear} solution~method.}.
As a nice feature of this model, the condition that two adjacent vertices
cannot share a common layer can simply be expressed as the variable
fixings (\ref{con:noshareD}), i.e., the variables may just be omitted
in practice. Without accounting for these, the total
number of constraints is $2\lvert A \rvert \cdot Y + \lvert V \rvert$, and
the total number of variables is $\lvert V \rvert \cdot Y + \lvert A \rvert \cdot (Y-1)^2$.

In terms of the latter, the ordering-based CGL models
(a compacted reformulation of those in
\cite{JabrayilovMMR+16} and \cite{JGAA-441}
is displayed in the appendix)
are more economical. Even if auxiliary
variables for arc reversals or dummy vertices are introduced,
their total number is still only $(\lvert V \rvert + \lvert A \rvert) \cdot (Y-1)$.
However, the CGL models induce about twice the number of
constraints compared to the model above (which is more critical
to a MIP solver), and they are more sensitive to a graph's~density.

For any arc $uv \in A$, there is exactly one pair of layers $k$ and $\ell$,
$k \neq \ell$, such that $x_{u,k} \cdot x_{v,\ell} = 1$. All other
products are zero. The length of $uv \in A$ thus
equals
    \[    \tssl_{\mathclap{k=2}}^{\mathclap{Y}} \tssl_{\mathclap{\;\; \ell=1}}^{\mathclap{\; k-1}}\; \left( (k - \ell) \cdot (x_{u,\ell} \cdot x_{v,k} + x_{u,k} \cdot x_{v,\ell}) \right) =
    \tssl_{\mathclap{k=2}}^{\mathclap{Y}} \tssl_{\mathclap{\;\; \ell=1}}^{\mathclap{\; k-1}}\; \left( (k - \ell) \cdot (p_{u,\ell,v,k} + p_{u,k,v,\ell}) \right).\]

Similarly, the arc $uv \in A$ is reversed
if and only if the expression
    \[\tssl_{k=2}^Y \left( x_{u,k} \cdot \tssl_{\ell=1}^{k-1} x_{v,\ell}\right) =
    \tssl_{k=2}^Y \tssl_{\ell=1}^{k-1} p_{u,k,v,\ell}\]
evaluates to one. Otherwise, the expression will evaluate to zero.

Finally, an arc $uv \in A$ causes a dummy vertex on a layer $k \in \{2, \dots, Y-1\}$
if and only if $k$ is between the layers $u$ and $v$ are assigned to, i.e., if
the expression
    \[
\tssl_{\ell=1}^{k-1} \tssl_{m=k+1}^Y (p_{u,\ell,v,m} + p_{u,m,v,\ell})
    \]
evaluates to one. Again, it will be zero otherwise.

\subsection{Quadratic Layer Assignment for GLP-W (QLA-W)}

To build model QLA-W from the above basis, it suffices to add
a additional continuous variable $\mathcal{W} \in \mathbb{R}_{\ge 0}$ to capture the width
together with the $Y$ constraints:
\begin{align*}
	   & \tssl_{\mathclap{v \in V}} x_{v,k}    &\le\; & \mathcal{W}          && \mbox{for all } k \in \{1, Y\}\\ 
	   & \tssl_{\mathclap{uv \in A}}\;\; \tssl_{\mathclap{\;\ell=1}}^{\mathclap{\;k-1}} \tssl_{\mathclap{\;\;\;\;\;\;\;\;\;m=k+1}}^{\;Y} (p_{u,\ell,v,m} + p_{u,m,v,\ell}) +
	    \tssl_{\mathclap{v \in V}} x_{v,k}     &\le\; & \mathcal{W} && \mbox{for all } k \in \{2, \dots, Y-1\} 
\end{align*}
The objective function can be stated as
    \[   \mbox{minimize }  \tssl_{\mathclap{uv \in A}}\ \big(
    \tssl_{\mathclap{k=2}}^{\mathclap{Y}} \tssl_{\mathclap{\;\; \ell=1}}^{\mathclap{\;\; k-1}}
     \omega_{len} (k - \ell) (p_{u,\ell,v,k} + p_{u,k,v,\ell}) +
     \omega_{rev}\; p_{u,k,v,\ell} \big) + \omega_{wid}\; \mathcal{W}.  \]

\subsection{Quadratic Layer Assignment for GLP-MS$^*$ (QLA-MS$^*$)}\label{s:QLA-MS}

Recalling the definitions from Sect.~\ref{s:Landscape},
GLP-MS$^*$ asks for the (weighted) minimization of
the inverse scaling factor
\begin{align*}
\bar{\mathcal{S}} = \frac{1}{\mathcal{S}} = \frac{1}{\min \{ \frac{r_W}{\mathcal{W}}, \frac{r_H}{\mathcal{H}} \}} = \max \big\{ \frac{\mathcal{W}}{r_W}, \frac{\mathcal{H}}{r_H} \big\}.
\end{align*}

Thus, to obtain model QLA-MS$^*$, the basic model is to be
extended with an according variable
$\bar{\mathcal{S}} \in \mathbb{R}_{\ge 0}$, and
with the $Y + \lvert V \rvert$ constraints:
\begin{align*}
	   & \tssl_{\mathclap{v \in V}} x_{v,k}    &\le\; & r_W\; \bar{\mathcal{S}}          && \mbox{for all } k \in \{1, Y\} \\
	   & \tssl_{\mathclap{uv \in A}}\;\; \tssl_{\mathclap{\;\ell=1}}^{\mathclap{\;k-1}} \tssl_{\mathclap{\;\;\;\;\;\;\;\;\;m=k+1}}^{\;Y} (p_{u,\ell,v,m} + p_{u,m,v,\ell}) +
	    \tssl_{\mathclap{v \in V}} x_{v,k} \!\!\!     &\le\; & r_W\; \bar{\mathcal{S}} && \mbox{for all } k \in \{2, \dots, Y-1\} \\
    & \tssl_{k=1}^{Y} k\; x_{v,k}           &\le\; & r_H\; \bar{\mathcal{S}} && \mbox{for all } v \in V
\end{align*}
Finally, the objective function is
    \[
 \mbox{minimize }  \tssl_{\mathclap{uv \in A}}\ \big(
    \tssl_{\mathclap{k=2}}^{\mathclap{Y}} \tssl_{\mathclap{\;\; \ell=1}}^{\mathclap{\;\; k-1}}
     \omega_{len} (k - \ell) (p_{u,\ell,v,k} + p_{u,k,v,\ell}) +
     \omega_{rev}\; p_{u,k,v,\ell} \big) + \omega_{scl}\; \bar{\mathcal{S}}.  \]

\setlength{\tabcolsep}{4pt}

\section{Experimental Evaluation}\label{s:Exp}

The \emph{aesthetic effects}
when integrating GLP-MS and GLP-W
into the hierarchical framework by
Sugiyama et al.~were extensively studied
already in~\cite{JabrayilovMMR+16,RueeggESvH15,RueeggESvH16,JGAA-441}.
Here, we thus confine ourselves to evaluate (i)
\emph{the computational effects when targeting
different aesthetic objectives and aspect ratios}, and (ii)
\emph{the competitiveness of QLA-W and QLA-MS$^*$
with respect to CGL-W and CGL-MS$^*$}.

To accomplish this, we employed the original models
CGL-W from~\cite{JabrayilovMMR+16} and
CGL-MS$^*$~from~\cite{JGAA-441}, except for leaving out
constraints that enforce at least one vertex to be
placed on the first layer\footnote{
Any solution violating these
constraints may be normalized a posteriori by simply ignoring empty layers.
Moreover, in case of GLP-MS$^*$, they are implied if the height
imposes a stronger restriction on the minimization of
$\bar{\mathcal{S}}$ than the width (which depends on the adjacency
structure of the graph as well as on the choice of $r_W$, $r_H$, and~$Y$).
In any other case, they do break some symmetries, but did not lead to
better experimental results.}.
Also, we employ the same two instance sets as in the mentioned prior
studies: The first set \textsc{ATTar} are the AT\&T graphs
from~\cite{DiBattistaGLP+97} whereof we extracted all non-tree
instances with
$20 \le \lvert V \rvert \le 60$, and $20 \le \lvert A \rvert \le 168$.
Their density $\frac{\lvert A \rvert}{\lvert V \rvert}$ is within 
$[1, 4.72]$ (on average $1.47$).
The second set \textsc{Random} consists of $180$ randomly generated
and also acyclic and non-tree graphs with $17$ to $60$ vertices,
and $30$ to $91$ arcs.
These were obtained as follows: First, a respective number of
vertices was created. Then, for each vertex, a random number of
outgoing arcs (with arbitrary random target) is added such that
the total number of arcs is $1.5$ times the number of vertices.
Finally, isolated vertices were removed.

During our experiments, all MIPs were solved using
Gurobi\footnote{A proprietary MIP solver, see \url{https://www.gurobi.com}}
(release version 8) single-threadedly on a Debian Linux system
with an Intel Core i7-3770T CPU ($2.5$~GHz)
and $8$~GB~RAM, and with a time limit set to half an hour.

\subsection{GLP and GLP-W}

Here, we consider two experiments. In experiment (1),
we set $Y = \lceil 1.6 \cdot \sqrt{\lvert V \rvert} \rceil$, $\omega_{len} = 1$,
$\omega_{rev} = Y \omega_{len} \lvert A \rvert$, and $\omega_{wid} = 1$
as in~\cite{JabrayilovMMR+16}. This can be seen as an almost
pure GLP setting, since one saved unit of width has the same (small)
effect in the objective function as one saved unit of (total) edge length.
Moreover, arcs are reversed only if this is inevitable due to the
choice of $Y$ or because they are part of a cycle. In experiment (2),
we instead give priority to width minimization by
increasing $\omega_{wid}$ to
$\omega_{rev} \lvert A \rvert + \lvert A \rvert \cdot Y  + 1$.

\begin{figure}[tbh]
\centering
    \resizebox{0.325\columnwidth}{!}{\input{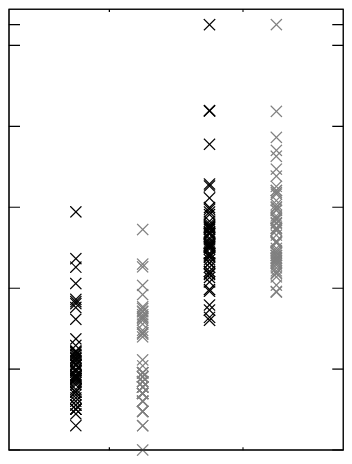}}
    \resizebox{0.325\columnwidth}{!}{\input{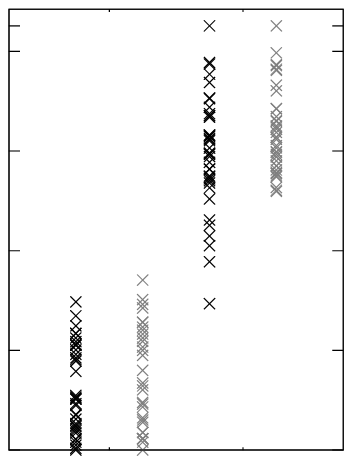}}
    \resizebox{0.325\columnwidth}{!}{\input{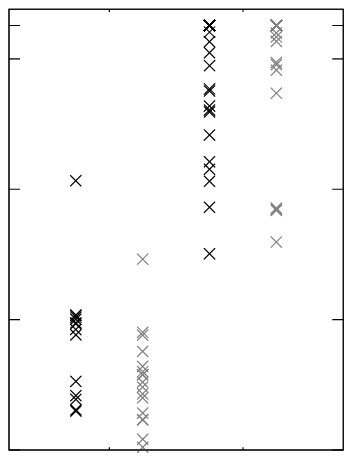}}
    \resizebox{0.325\columnwidth}{!}{\input{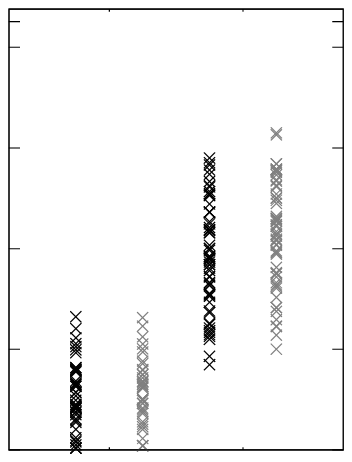}}
    \resizebox{0.325\columnwidth}{!}{\input{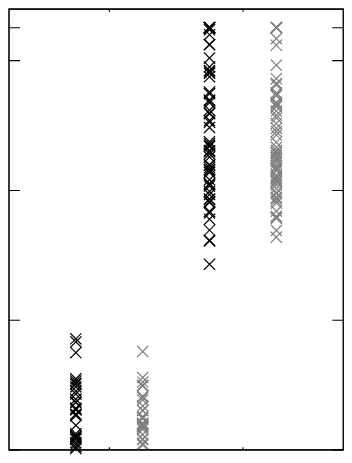}}
    \resizebox{0.325\columnwidth}{!}{\input{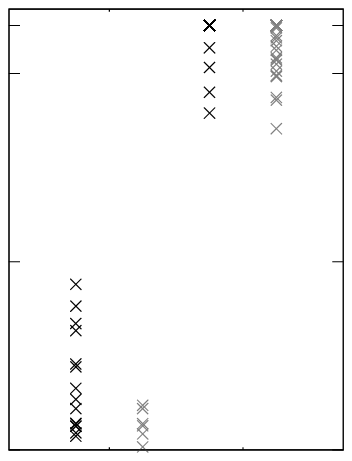}}
    \resizebox{0.42\columnwidth}{!}{\input{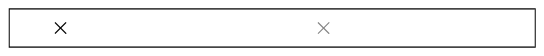}}
    \caption{The plots depict the solution times in seconds (a cross per graph) for
    GLP with neglected (1) and emphasized width minimization (2).
    Crosses at the top (\mbox{1800$s$}) correspond to timeouts, their quantities are given to the left
    of the respective cross.}
\label{fig:plotW}
\end{figure}

The results are shown in Fig.~\ref{fig:plotW}. The most distinctive
observation is that the layering problem is considerably harder to solve with
\emph{both} MIPs if emphasis is given to width minimization. In this case,
the solution times are higher and several timeouts occur for the
larger graphs, while the pure GLP setting can be solved routinely for
all instances considered. With respect to experiment (2), the results
obtained with QLA-W are slightly better on average than with CGL-W
for the \textsc{ATTar} graphs, whereas the opposite is true for the
\textsc{Random} graphs, especially due to the increased number of timeouts
for the larger ones. Thus, in total, there is no clear superior or
inferior model -- on average both show a comparable or competitive
performance with the MIP solver employed.

\subsection{GLP-MS$^*$}

In this experiment, the parameters are set (almost) as
in~\cite{JGAA-441}: $Y = \lvert V \rvert$, $\omega_{len} = 1$,
$\omega_{rev} = Y \omega_{len} \lvert A \rvert$, and
$\omega_{scl} = \omega_{rev} \lvert A \rvert + \lvert A \rvert \cdot Y  + 1$.
Priorities are thus the same as in experiment (2) before, except
now emphasizing on a maximum scaling factor instead of the width alone.
The ${r_H:r_W}$ ratios considered are ${1:2}$, ${1:1}$, and~${2:1}$.

\begin{figure}[tbh]
\centering
    \resizebox{0.325\columnwidth}{!}{\input{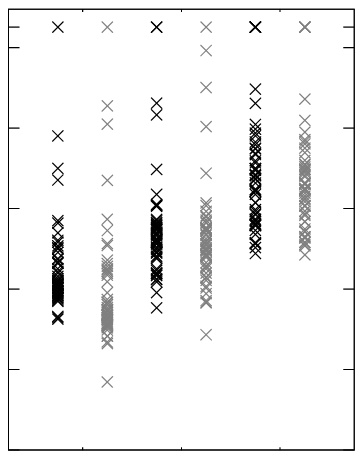}}
    \resizebox{0.325\columnwidth}{!}{\input{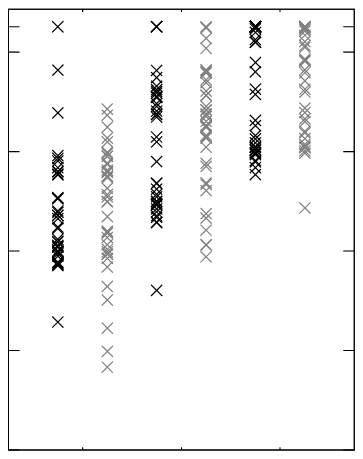}}
    \resizebox{0.325\columnwidth}{!}{\input{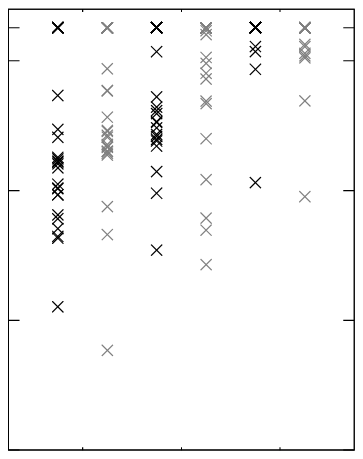}}
    \resizebox{0.325\columnwidth}{!}{\input{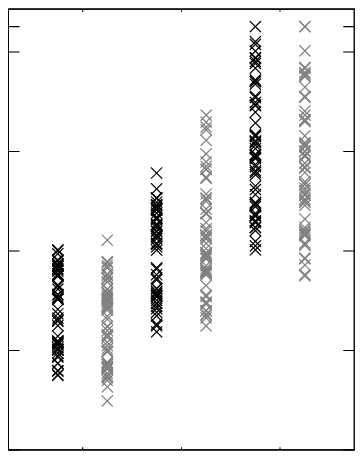}}
    \resizebox{0.325\columnwidth}{!}{\input{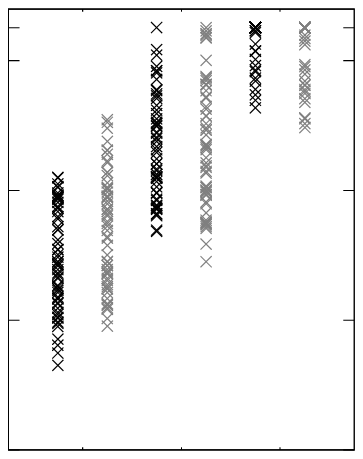}}
    \resizebox{0.325\columnwidth}{!}{\input{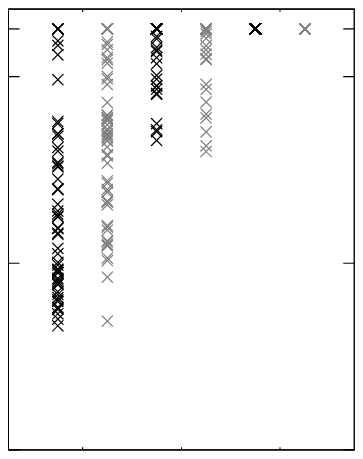}}
    \resizebox{0.40\columnwidth}{!}{\input{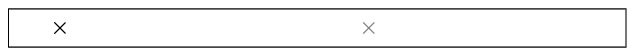}}
    \caption{The plots depict the solution times in seconds (a cross per graph) for the different ${r_H:r_W}$ combinations.
    Crosses at the top (\mbox{1800$s$}) correspond to timeouts, their quantities are given to the left
    of the respective cross.}
\label{fig:plotMS}
\end{figure}

As Fig.~\ref{fig:plotMS} shows, the results are again
diverse under fine-grained inspection, and comparable
from a coarse perspective. For both QLA-MS$^*$ and CGL-MS$^*$
there are instance subsets and ${r_H:r_W}$ ratios where the
use of either one leads to faster solutions and less timeouts.

Also, for both formulations, the ${2:1}$-case, where the height
is constrained to be \emph{at most} twice the width, turns out
to be the hardest - which is not surprising as e.g.~the
\textsc{ATTar} graphs are \emph{at least}
twice as high as wide when drawn with the classic framework~\cite{JabrayilovMMR+16}.
As opposed to that, in the ${1:2}$-case, the condition that
the height is restricted to be no more than
half of the width is a strong one, as most of
the considered graphs cannot be layered very widely.
This entails better bounds on the objective function
during the MIP solution process.

An interesting question is the relative tractability
of GLP-W and (especially the ${2:1}$-case of) the more
general GLP-MS$^*$, since (in a sense)
both aim at a (relatively) small width, but the first with respect
to a fixed (but larger) height limit, and the second additionally
requiring to find the best ${\mathcal{H}:\mathcal{W}}$-pair
under the given aspect ratio constraint. As could be expected,
comparing the two results indicates that GLP-W appears to be
considerably easier to solve to optimality, at least with
the present modeling techniques.

\section{Conclusion}\label{s:Concl}

In recent years, several extensions of the graph layering problem
of increasing generality have been studied, and several
algorithmic solution approaches have been proposed. Exact methods
are based on integer programming where assignment and ordering
variables have dominated the most successful models for the more
general variants of the problem.

In this paper, we proposed a
new \emph{quadratic} assignment approach that can be adapted
to each of these, allows a natural expression of the associated
layout restrictions or aesthetic objectives, and turns out to
be computationally competitive. However, as soon as the emphasis
is on width minimization or even a particular drawing area is
targeted, still neither of the present methods is able to
\emph{routinely} solve \emph{arbitrary} instances beyond about
$50$ vertices. Moreover, as long as the proposed quadratic model
is solved based on linearization, its problem size will become
a limitation if the number of vertices is considerably increased.
Nevertheless, depending on the adjacency structure
of the graphs to be layered, this may be possible, and our results
show that different settings (aspect ratios, emphasis or neglection of
the width, restrictive or non-restrictive heights)
have a strong impact on the tractability of the problem -- which
also means that some can be handled quite well.

Also, the recent generalizations
of the layering problem are a clear step towards the requirements
of real-world applications -- and real-world displays. We
hope that our drawn links to the quadratic assignment problem
inspire also some novel heuristic solution approaches, or exact
models for related graph drawing paradigms.

\appendix

\section{Ordering-based Reference Models}\label{s:RefMIP}

As a reference, we display here slightly more compact reformulations
of CGL-W and CGL-MS$^*$ compared to their respective original presentations
in~\cite{JabrayilovMMR+16} and~\cite{JGAA-441}.

Let $G=(V,A)$ be a directed graph, and let $Y$ be an upper bound
on the height of the layering to be chosen a priori. A basic
CGL model then involves binary variables $y_{k,v}$, for all
${v \in V}$ and all layer indices $k \in \{1, \dots, Y-1\}$,
being equal to $1$ if $k < L(v)$ and being equal to zero
otherwise. In particular $L(v) = 1$ if and only if $y_{1,v} = 0$,
$L(v) = Y$ if and only if $y_{k,v} =1$ for all
$k \in \{1, \dots, Y-1\}$, and $L(v) = k$ if and
only if $y_{k-1,v} - y_{k,v} = 1$.
The number of basic variables thus amounts to $\lvert V \rvert \cdot (Y-1)$.

Moreover, there are only the following $\lvert V \rvert \cdot (Y-2)$ basic constraints 
that establish transitivity in the sense that
$L(v) > k$ implies $L(v) > k - 1$ for each $k \in \{2, \dots, Y-1\}$:
\begin{align*}
	   &  y_{k,v} - y_{k-1,v}                 &\le\; &  0         && \mbox{for all } v \in V,\; k \in \{2, \dots, Y-1\} 
\end{align*}

However, to model GLP, $\lvert A\rvert$ further auxiliary variables
$r_{uv}$ as well as $\lvert A \rvert  \cdot Y$ constraints -- ensuring that
$r_{uv}$ is equal to one if $uv \in A$ is reversed and equal to zero otherwise --
are required.

These constraints are:
\begin{align*}
           &  y_{1,u}             - r_{uv}        &\ge\; &  0         && \mbox{for all } uv \in A  \\
           &  y_{1,v}             + r_{uv}        &\ge\; &  1         && \mbox{for all } uv \in A \\
           &  y_{k-1,u} - y_{k,v} - r_{uv}        &\le\; &  0         && \mbox{for all } uv \in A,\; k \in \{2, \dots, Y-1\} \\
           &  y_{k-1,v} - y_{k,u} + r_{uv}        &\le\; &  1         && \mbox{for all } uv \in A,\; k \in \{2, \dots, Y-1\}\\
           &  y_{Y-1,u}           - r_{uv}        &\le\; &  0         && \mbox{for all } uv \in A \\
           &  y_{Y-1,v}           + r_{uv}        &\le\; &  1         && \mbox{for all } uv \in A 
\end{align*}

Moreover, to model dummy vertices (and thus the width),
$\lvert A \rvert \cdot (Y-2)$ further variables $d_{uv,k}$
for each arc $uv \in A$ and each layer
$k \in \{2, \dots, Y-1\}$ are required. The associated
constraints enforcing $d_{uv,k}$ to be one if $uv \in A$
causes a dummy vertex on layer $k$
(otherwise, an optimum solution has $d_{uv,k} = 0$ whenever $\omega_{len} > 0$)~are:
\begin{align*}
           &  y_{k,u} - y_{k-1,v} - d_{uv,k}      &\le\; &  0         && \mbox{for all } uv \in A,\; k \in \{2, \dots, Y-1\} \\
           &  y_{k,v} - y_{k-1,u} - d_{uv,k}      &\le\; &  0         && \mbox{for all } uv \in A,\; k \in \{2, \dots, Y-1\}\\
\end{align*}

To obtain CGL-W, one further adds a variable $\mathcal{W}$ and the $Y$ constraints:
\begin{align}
     & \tssl_{\mathclap{v \in V}} ( 1 - y_{1,v} ) &\le\; & \mathcal{W}&&    \label{cgl:width1} \\
     & \tssl_{\mathclap{v \in V}}     y_{Y-1,v}   &\le\; & \mathcal{W}&&    \label{cgl:widthY} \\
     & \tssl_{\mathclap{v \in V}} ( y_{k-1,v} - y_{k,v} ) +
       \tssl_{\mathclap{uv \in A}} d_{uv, k}      &\le\; & \mathcal{W}     && \mbox{for all } k \in \{2, \dots, Y-1\} \label{cgl:width}
\end{align}

The total number of constraints is then $(4\lvert A \rvert + \lvert V \rvert + 1) \cdot (Y-2) + 4 \lvert A \rvert + 2$.
One can now exploit that
the length of an arc (i.e., the difference of the layer
indices its endpoints are assigned to) is equivalent
to the number of dummy vertices it causes plus one.
Thus, the objective function for CGL-W can be expressed as:
    \[
	\mbox{minimize } \tssl_{\mathclap{uv \in A}} \big( \omega_{rev}\; r_{uv} + \omega_{len} (1 + \! \tssl_{k=2}^{Y-1} \! d_{uv,k}) \big) +  \omega_{wid}\; \mathcal{W} \]

To rather obtain CGL-MS$^*$, it suffices to
introduce the variable $\bar{\mathcal{S}}$ instead, replace $\mathcal{W}$ with
$r_W\; \bar{\mathcal{S}}$ in (\ref{cgl:width1})--(\ref{cgl:width}), and to add
the $\lvert V \rvert$ additional constraints:
\begin{align*}
      & 1 + \tssl_{\mathclap{k \in \{1, \dots, Y-1\}}} y_{k,v} &\le\; & r_H\; \bar{\mathcal{S}} && \mbox{for all } v \in V  
\end{align*}

Then, the objective is
\[
    \mbox{minimize } \tssl_{\mathclap{uv \in A}} \big( \omega_{rev}\; r_{uv} + \omega_{len} (1 + \! \tssl_{k=2}^{Y-1} \! d_{uv,k}) \big) + \omega_{scl}\; \bar{\mathcal{S}}.\]
and the total number of constraints amounts to $(4\lvert A \rvert + \lvert V \rvert + 1) \cdot (Y-2) + 4 \lvert A \rvert + \lvert V \rvert + 2$.

\bibliographystyle{splncs04}
\bibliography{references}

\end{document}